\documentclass{JINST_paper}
\usepackage{subfigure}
\usepackage{amsmath}
\usepackage{booktabs}
\usepackage{array} 

\title{Design, Implementation and First Measurements with the Medipix Neutron Camera in CMS}

\author{
Austin Ball$^b$,
Alan Bell$^b$, 
Anthony Butler$^a$, 
Philip Butler$^a$, 
Richard Hall-Wilton$^b$\thanks{now at European Spallation Source AB.}$\ $, 
Jeroen Hegeman$^b$,
Stuart Lansley$^a$,  
Gary Keen$^c$,
David Krofcheck$^e$,
Steffen Mueller$^b$,
Alick Macpherson$^b$,
Dorothea Pfeiffer$^a$\thanks{Corresponding
author.}$\ $,
Stanislav Pospisil$^d$,
Hamish~Silverwood$^a$,
Emmanuel Tsesmelis$^b$,
Zdenek Vykydal$^d$
\\
\llap{$^a$}University of Canterbury\\
  Christchurch, New Zealand\\
\llap{$^b$}CERN\\
  CH-1211 Geneva 23, Switzerland\\
\llap{$^c$}Rensselaer Polytechnic Institute\\
Troy (NY), USA\\
\llap{$^d$}Institute of Experimental and Applied Physics (IEAP) - Czech Technical University (CTU)\\
Prague, Czech Republic\\
\llap{$^e$} University of Auckland\\
Auckland, New Zealand\\
  E-mail: \email{Dorothea.Pfeiffer@cern.ch}}

\abstract{
The Medipix detector is the first device dedicated to measuring mixed-field radiation in the CMS cavern and able to distinguish between different particle types. Medipix2-MXR chips bump bonded to silicon sensors with various neutron conversion layers developed by the IEAP CTU in Prague were successfully installed for the 2008 LHC start-up in the CMS experimental and services caverns to measure the flux of various particle types, in particular neutrons. They have operated almost continuously during the 2010 run period, and the results shown here are from the proton run between the beginning of July and the end of October 2010. Clear signals are seen and different particle types have been observed during regular LHC luminosity running, and an agreement in the measured flux rate is found with the simulations.
These initial results are promising, and indicate that these devices have the potential for further and future LHC and high energy physics applications as radiation monitoring devices for mixed field environments, including neutron flux monitoring. Further extensions are foreseen in the near future to increase the performance of the detector and its coverage for monitoring in CMS.
}


\begin{document}

\section{Introduction}
\begin{figure}[htbp]
  \centering
\subfigure[Medipix detector]{
	\includegraphics[width=8 cm]{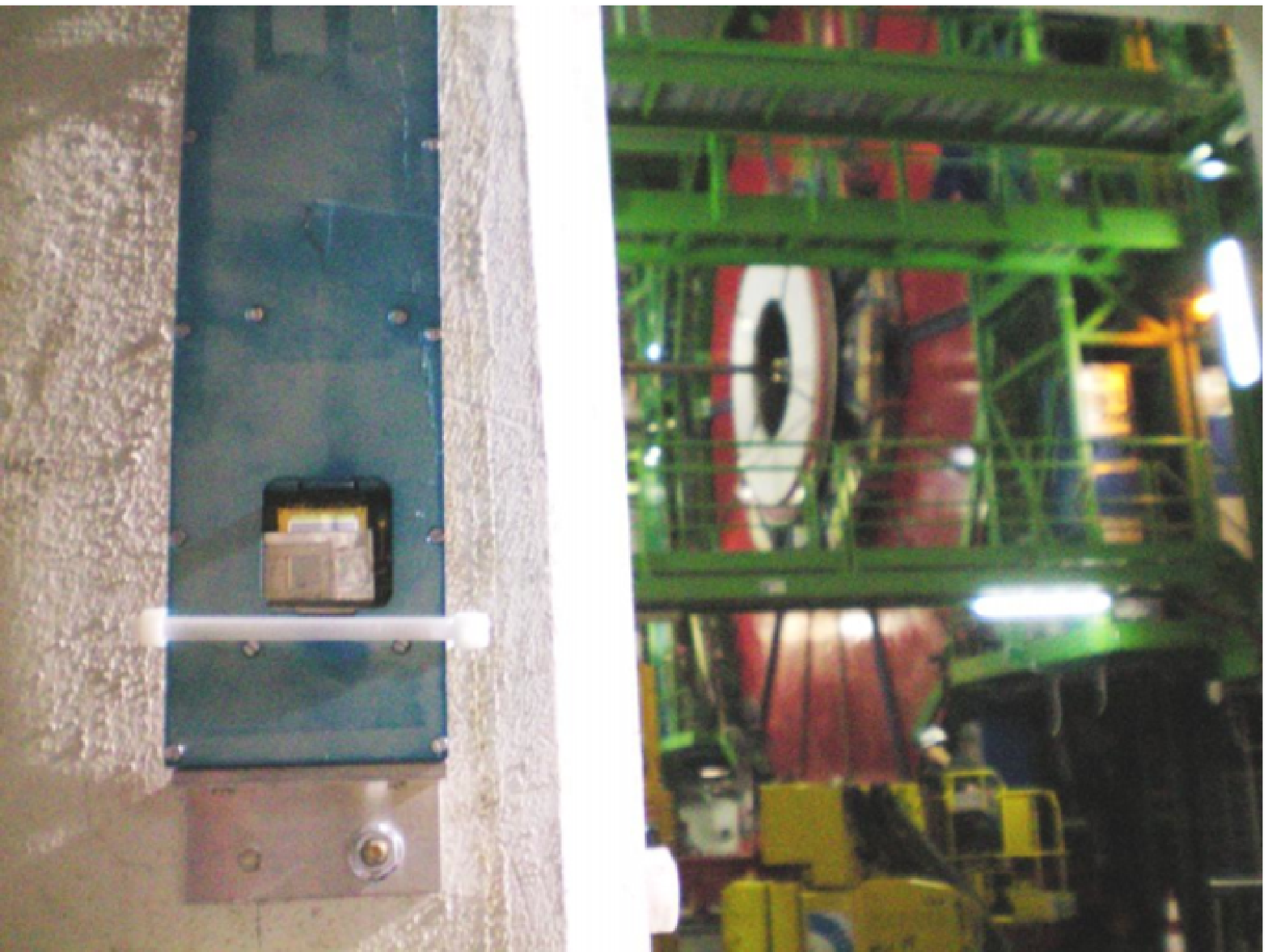}
	\label{fig: Medipix detectors}
}
\hfill
\subfigure[Neutron conversion layers]{
	\includegraphics[width=6 cm]{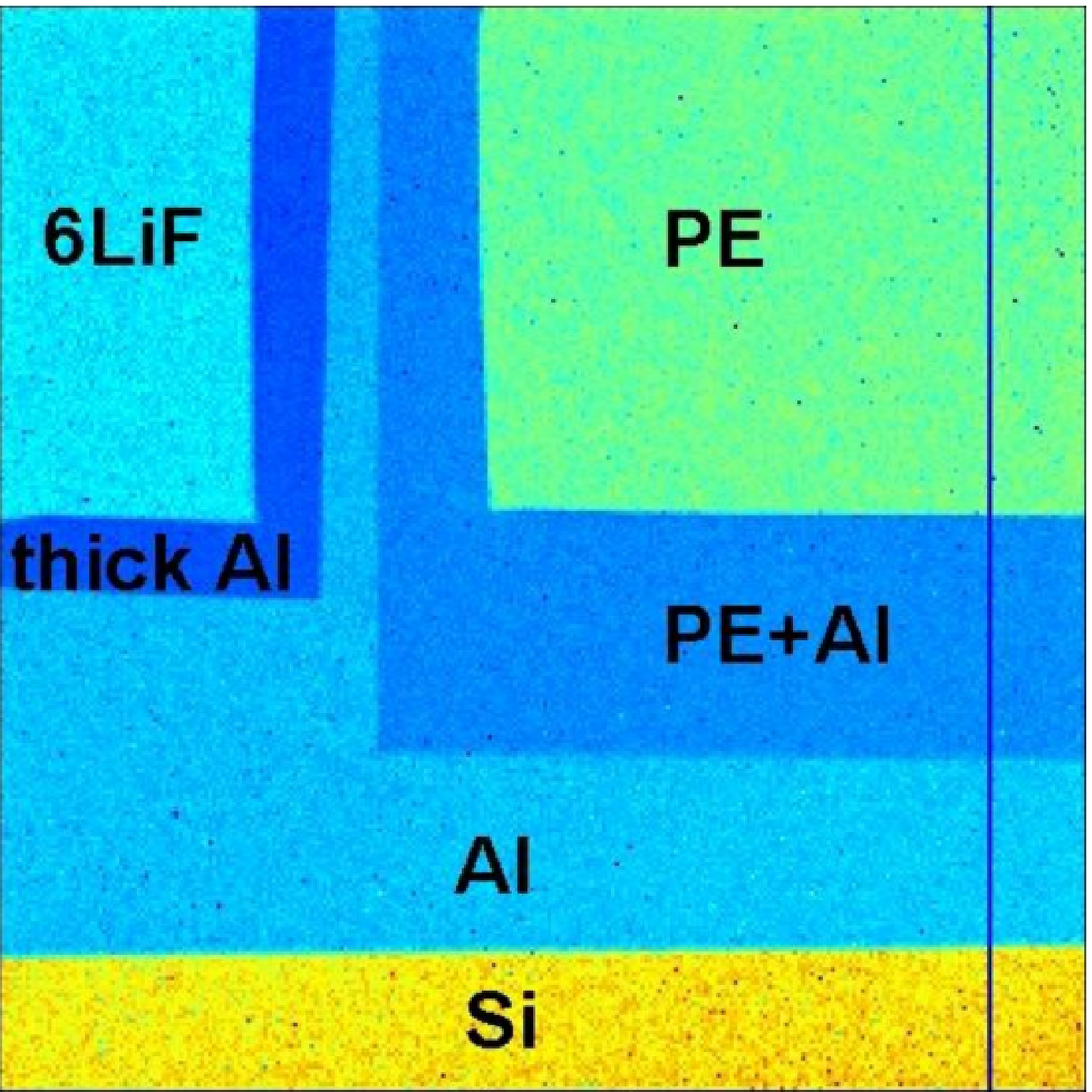}
	\label{fig: Conversion layers}
}
\caption{Example of a Medipix detector installed in the CMS underground area (a) and neutron conversion layers attached to the silicon sensor on the Medipix2-MXR chip of the detector in the CMS cavern (b).}
\label{fig: Medipix detectors and conversion layers}
\end{figure}

Three Medipix2 detectors were installed in the CMS \cite{CMS} experiment at the LHC \cite{LHC} as part of the CMS beam and radiation monitoring system \cite{Chong:2007,Fernandez:2005,Macpherson:2006,BRM,BCM1F} in March 2008 for the LHC start-up. An example of the detector is shown in figure 
\ref{fig: Medipix detectors}. The motivation for the Medipix project in CMS was to show that different types of particles like neutrons, photons and electrons can be detected in the cavern, similar to the aim of the ATLAS-MPX project \cite{ATLAS, Campbell:2007, Vykydal:2008a}. Further, it was the goal to determine whether a correlation between the luminosity and the number of detected particles exists and then compare the measured fluxes with the results from simulations. In this sense the Medipix project was intended as proof of principle project with the aim to extend it if the detectors gave promising results. 

Due to magnetic field and cable length restrictions of the readout, the present detectors had to be installed in particular locations, i.e. it was not possible to install them close to the beam line. The interaction point for proton-proton collisions at CMS is x = 0 m, y = 0 m and z = 0 m. One detector was installed at position x = 11.5 m, y = -2 m and z = 12 m in the CMS cavern, one just outside the mobile shielding of the cavern and one in the underground services cavern S1 (x = 20 m, y = 2 m, z = 0 m) at Point 5 of the LHC. At the same locations CERN RADMON \cite{Pignard,Wijnands:2005} systems are installed. Unfortunately, the detector outside of the mobile shielding was damaged several times by leaking water and then destroyed in June 2010, so that only two detectors were operational and used for the study presented in this paper. 

\section{The Medipix2-MXR detector and neutron conversion layers}

\subsection{Medipix2-MXR}
The Medipix2-MXR chip \cite{Llopart:2002} was developed by the Medipix collaboration \cite{MedipixCollaboration, Campbell:2010}, and contains 256 $\times$ 256 pixels with an edge length of 55 $\mu$m. At CMS, the Medipix2-MXR chip bump bonded to a 300 $\mu$m thick silicon sensor was used in combination with a USB 1.1 readout \cite{Vykydal:2006}. Due to cable lengths of more than 5 meters a USB to Ethernet extender had to be used, which required an additional power supply. Measurements with the Medipix2-MXR can be fine-tuned by setting certain parameters or DAC values. Among those are a coarse and fine lower energy threshold THL and a coarse and fine higher energy threshold THH. For all measurements at CMS and in the laboratory Pixelman \cite{Holy:2006}, a software package for Medipix detectors, has been used to set the measurement parameters and to acquire the data.

\subsection{Neutron conversion layers}
\label{Neutron Conversion Layers}
The IEAP CTU in Prague has extended the kinds of detectable particles by adding layers on top of the silicon sensor for neutron conversion and supplementary discriminative power between particle types \cite{Jakubek:2006,Uher:2005}. All chips used at CMS had identical conversion layers with slightly different dimensions on each detector. The layer sizes of the sensor in the CMS cavern can be found in table \ref{table: Particle track fluxes}, and a picture of the layers is shown in figure \ref{fig: Conversion layers}. Whereas one strip at the bottom of the detector is left uncovered, the rest of the surface is covered by layers. $\mathrm{^{6}{Li}}$ in the form of $\mathrm{^{6}{LiF}}$ powder sputtered to the bottom of a 50 $\mu$m thick aluminium foil covers the upper left corner and creates a layer aimed at the conversion of thermal neutrons (energy < 100 ke$\,$V) into $\mathrm{\alpha}$-particles via the reaction $\mathrm{^{6}Li\, + \,n \,\rightarrow \,\alpha\, + \,^3H}$. The yield of this reaction decreases with increasing energy of the incoming neutrons. The polyethylene (PE) layer is a rectangular block of 1.3 mm thickness and is glued to the upper right corner of the sensor. PE can convert fast neutrons into recoiled protons following the reaction equation $\mathrm{H\, +\, n\, \rightarrow \, p\, + \,n}$. Up to an energy of 15 Me$\,$V of the incoming neutrons the yield of this reaction increases and then decreases for higher energies. Fast neutrons can also interact directly with silicon and produce an $\mathrm{\alpha}$-particle according to $\mathrm{^{28}Si \, + \, n \, \rightarrow \, \alpha \, + \, ^{25}Mg}$ or produce a proton via $\mathrm{^{28}Si \, + \, n \, \rightarrow \, p \, + \, ^{28}Al}$. Aluminium layers of 100 $\mu$m and 150 $\mu$m thickness are applied which slightly change the conversion efficiency by attenuating or absorbing recoiled protons. The efficiencies of these conversion reactions have been determined in \cite{Greiffenberg:2009, Vykydal:2008b, Uher:2008}. This calibration was carried out for each device used here to within a statistical uncertainty of 2$\%$.

\section{Simulation of radiation background in the CMS cavern}
The Medipix detector is the first device dedicated to measuring mixed-field radiation in the CMS cavern and able to distinguish between different particle types. Extensive simulations of the radiation background have been carried out \cite{Mueller:2010p, Mueller:2010t}. The results are shown in table \ref{table: Simulated fluxes} and provide a solid base of comparison for the Medipix results. All simulation results presented in this work were obtained with FLUKA \cite{Ferrari:2005zk,Battistoni:2007zzb} using an updated CMS geometry based on \cite{Huhtinen:1996uf}. The pp-collisions are created by DPMJET III \cite{Roesler:2000he}. For the pp-collisions the flux-maps for the CMS cavern and the particle energy spectra for several locations have been simulated. In order to estimate the expected particle flux at the Medipix location, a simulation with 14 Te$\,$V pp-collisions was calculated and the flux of various particle types could be obtained. Details about the simulation setup can be found in \cite{Mueller:2010t}. The FLUKA simulations used a radius of 11.7 m - 11.9 m and a z position of 11 m - 12 m, corresponding to the location of the Medipix detector in the CMS cavern. The simulation results for 14 Te$\,$V pp-collisions were scaled with a multiplicity factor of $\sqrt{7\ \mathrm{Te\,V}/14\ \mathrm{Te\,V}}$ to compute the fluxes for 7 Te$\,$V pp-collisions \cite{Mueller:2010t}.

\begin{table}[htbp]
\begin{footnotesize}
        \centering
        \begin{tabular}{  l  c l c}
           		\toprule
        Particle & Simulated Flux (14 Te$\,$V) & Simulated Flux (7 Te$\,$V)\\  
			  &	$\left[\frac{particles}{cm^{2} \; s}/\frac{10^{30}}{ cm^{2}\; s}\right]$ &  
			  $\left[\frac{particles}{cm^{2} \; s}/\frac{10^{30}}{ cm^{2}\; s}\right]$
\\    
			\midrule
neutrons   (< 100 ke$\,$V)        & 0.1438(20) &  0.1017(14) \\ 
neutrons   (100 ke$\,$V - 20 Me$\,$V) & 0.0932(10) & 0.0659(07)\\ 
neutrons   (> 20 Me$\,$V)   			& 0.0257(04) &  0.0181(03)\\ 
neutrons  (all) 						  & 0.2627(17) &  0.1858(12) \\ 
charged hadrons               & 0.000534(62) & 0.000378(44) \\
electrons   								  & 0.0033(02) & 0.0023(01)\\ 
photons      								  & 0.1915(27) &  0.1354(19) \\ 
all          								  & 0.4582(33) &  0.3240(23) \\ 
\bottomrule
\end{tabular}
\caption{Simulated fluxes obtained with FLUKA simulations at the location of the Medipix detector in the CMS cavern.}                      
\label{table: Simulated fluxes}  
\end{footnotesize}
\end{table}
The neutron energy spectrum at the location of the Medipix detector in the CMS cavern is roughly constant as a first approximation. The majority of photons there arrive with an energy at which Compton scattering is the major interaction mechanism with silicon. Given the location of the Medipix detector on the CMS cavern wall, comparatively far away from the majority of the CMS detector material, an approximately normal incidence of particles can be assumed. This assumption is also supported by the simulations.

\section{Particle identification and detection efficiency}
\label{Particle identification and detection efficiency}
The first tests involved measurements of sources with well-defined radiation emissions. An $\mathrm{^{55}Fe}$,  a $\mathrm{^{90}Sr}$ source and a $\mathrm{^{241}Am}$ source were used to gain a better understanding of the energy resolution of the Medipix detector and to study its ability to distinguish between different particle types. Hit patterns were used to develop a particle identification algorithm, similar to the strategy described in \cite{Holy:2008}. The particle identification strategy adopted was based upon two discriminative techniques: firstly, differences in particle fluxes between the conversion layers due to the additional processes in these layers as indicated in section \ref{Neutron Conversion Layers}; secondly, differences in cluster sizes and track shapes caused by variations in energy deposition patterns as the particles pass through the Medipix detector. Starting from typical cluster shapes observed during the source measurements, a particle identification algorithm has been developed with the computer algebra system MATLAB \cite{MATLAB}, based upon existing categories and ROOT \cite{ROOT} code developed by the University of Montreal \cite{Idarraga:mafalda}. Nine different basic track shapes were defined and then identified in each frame as described in table \ref{table: Blob types created by different particle types}. 

Photons emitted by the $\mathrm{^{55}Fe}$ source (5.89 ke$\,$V) and the $\mathrm{^{241}Am}$ source (59 ke$\,$V) appear with a probability of more than 98$\%$ as ``single hits`` or ``double hits``, whereas less than 2$\%$ of the photons create a track from the group consisting of ``curly tracks``, ``MIPs`` (minimum-ionizing-particle-like shape), ``triple hits`` and ``quad hits``. Emitted $\mathrm{\alpha}$ particles produce almost exclusively ``heavy blobs``, less than 5$\%$ are identified as ``quadruple hits`` or ``curlies``. Emitted electrons generate the widest range of track shapes. About 40$\%$ of the electrons might be misidentified as photons because they appear as ``single hits``, ``double hits`` or ``long gammas``, and about 11$\%$ create a ``heavy blob`` or ``heavy track``. Nevertheless 49$\%$ correctly create electron-like clusters. The findings here are roughly comparable to those previously found by \cite{Bouchami:2010a}.

After combining the data gained from the sources, one arrives at table \ref{table: Blob types created by different particle types} which shows the three categories of cluster shapes: photon-like, electron-like and heavy-ionizing-particle-like shapes. The known probability of each particle type to create one element of the track categories makes it possible to determine the particle fluxes in the CMS cavern from the number of registered track shapes. 

\begin{table}[htbp]
\begin{footnotesize}
        \centering
        \begin{tabular}{  l c c c}
        \toprule
				Shapes 			 & \multicolumn{3}{c}{Particle Types} \\  				
              			& Photons  & Electrons  & Alpha Particles  \\ 
              		  & $[\%]$     &  $[\%]$     & $[\%]$           \\
				\midrule         															                             
        photon-like (``single hits``, ``double hits``, ``long gammas``)  &  99 		   &    40      &  0      \\ 
        electron-like (``curlies``, ``MIPs``, ``triple hits``, ``quad hits``)&   1       &    49      &  5      \\ 
        heavy-ionizing-particle-like (``heavy blobs``, ``heavy tracks``) &   0       &    11      & 95      \\ 
        \bottomrule
        \end{tabular}
        \caption{Shapes created by different particle types in the Medipix detector.}                      
        \label{table: Blob types created by different particle types}  
        \end{footnotesize}
\end{table}

For electrons in all layers a detection efficiency of 100$\%$ is assumed. Whereas during the source measurements of low energy photons the detection efficiency was above 50$\%$, the detection efficiency of photons in general is a lot lower and decreases for higher energy photons. 

Considering the absorption coefficient of photons in silicon and the simulated photon energy spectrum, the detection efficiency is approximately 1$\%$ for photons in the CMS cavern which have predominantly energies between 50 ke$\,$V and 500 ke$\,$V. The distinction between photons and electrons in the Medipix detector is generally difficult \cite{Teyssier:2010}. The probability of misidentifying particles depends on their incoming energy, their angle of incidence and the applied energy threshold for the measurement. Electrons can be misidentified as photons and vice versa. Additionally high energy photons also interact with other material in the cavern such as the CMS detector, and various absorption and conversion mechanisms are possible. On the other hand electrons are effectively filtered out at this location due to the CMS detector 3.8 T magnetic field, a factor which influences the detected admixture.

The only layer with a considerable conversion efficiency for thermal neutrons is the $\mathrm{^{6}{LiF}}$ layer. Using the efficiencies in \cite{Greiffenberg:2009, Vykydal:2008b, Uher:2008} and the approximately equal distribution of neutrons in the simulation between 10 eV and 100 ke$\,$V, one arrives at an average detection efficiency of around 0.5$\%$ for thermal neutrons in this energy range. 

For fast neutrons the PE layer has the largest detection efficiency \cite{Bouchami:2010b}. Combining the detection efficiencies of references \cite{Greiffenberg:2009, Vykydal:2008b} and applying them to the neutron spectrum results in an average detection efficiency of approximately 0.1$\%$ for neutrons interacting in the PE layer with energies between 100 ke$\,$V and 20 Me$\,$V. The detection efficiency for neutron interactions in the silicon itself increases with neutron energy between 0 Me$\,$V and 14 Me$\,$V. The simulation indicates that the spectrum of fast neutrons is relatively constant, thus a substantial proportion of neutrons have an energy above 14 Me$\,$V. Therefore the detection efficiency assumed for these neutron interactions in the silicon is taken as 0.1$\%$, which is the efficiency at the highest calibrated energy (14 Me$\,$V). The systematic uncertainties on these efficiencies due to the spectra in the CMS cavern, which differ from the spectra of the sources used for calibration, have yet not been evaluated in detail. Therefore the conversion efficiencies determined with calibration sources provide only an estimate regarding the conversion efficiencies for the Medipix2-MXR detector in the CMS cavern.

\section{Measurements during LHC operation}

After evaluating the source measurements, the parameters for the cavern data acquisition were determined. The measurements were made at a fixed lower energy threshold, and not while scanning successively over an energy range using lower and upper threshold. To detect also low energy particles a lower energy threshold of 15 ke$\,$V was chosen, a threshold that is low enough to detect almost all particles. A relatively long acquisition time of 60 s has been chosen to get reasonably high statistics per frame. Then the dead time for 60 s amounts to less than 5$\%$. A bias voltage of 100 V was applied for nominal detection efficiency.

\begin{figure}[htbp]
  \centering
\subfigure[linear scale]{
	\includegraphics[width=7.2 cm]{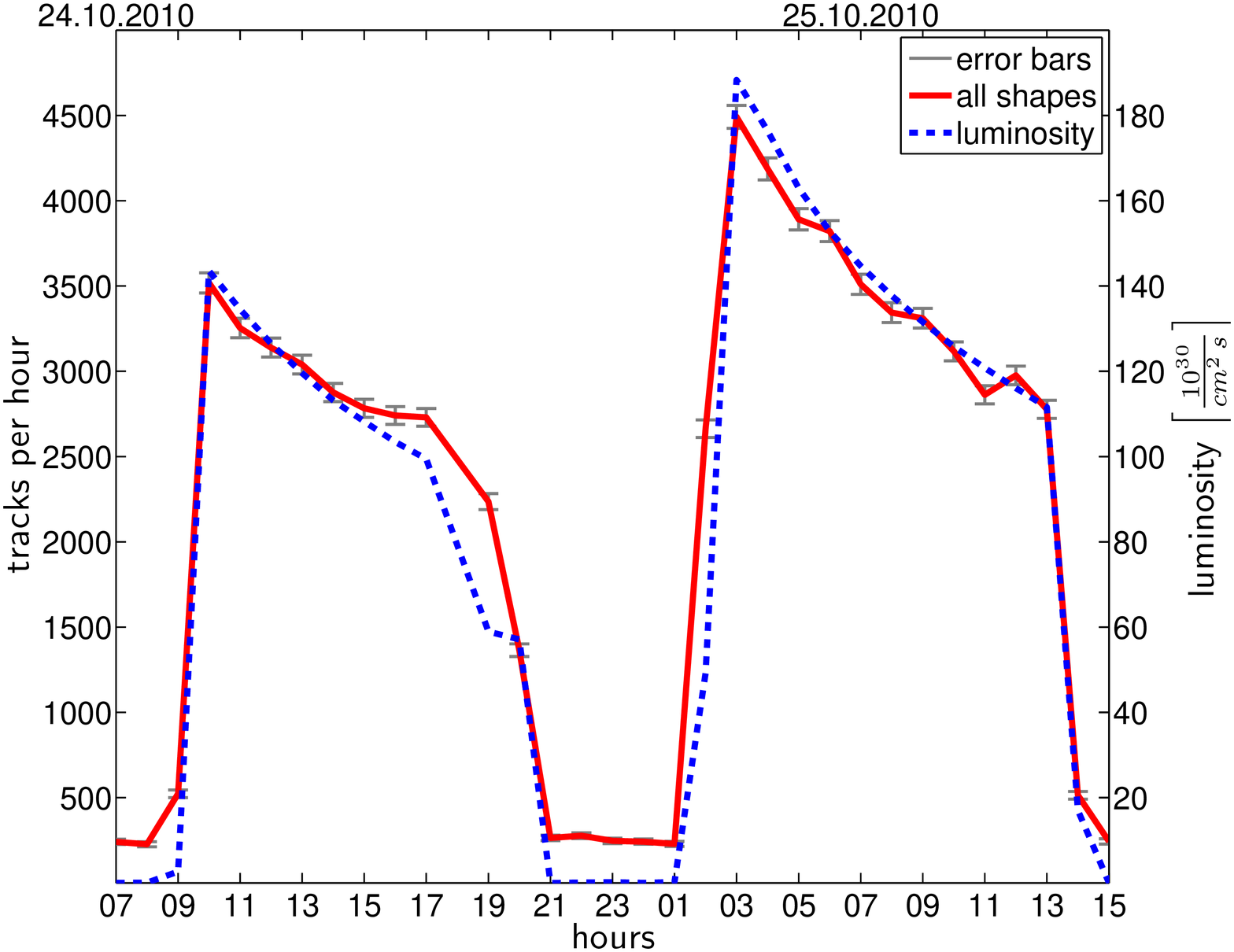}
	\label{fig: Cavern two fills linear scale}
}
\hfill
\subfigure[logarithmic scale]{
	\includegraphics[width=7.2 cm]{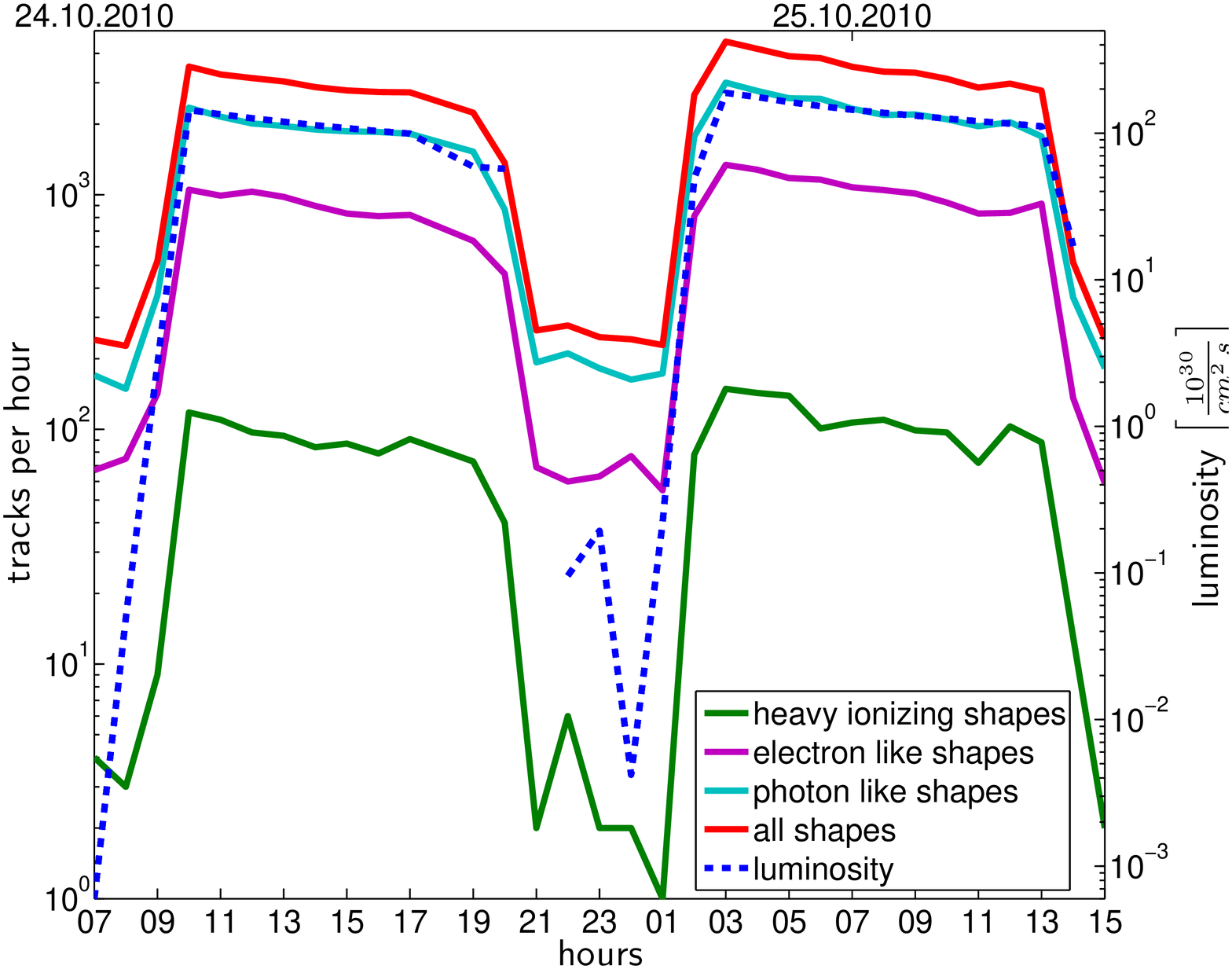}
	\label{fig: Cavern two fills logarithmic scale}
}
\caption{Analysis of instantaneous luminosity and registered track shapes in two complete fills as measured with the Medipix detector inside the CMS cavern.}
\label{fig: Cavern analysis complete fills}
\end{figure}

The data analysed in this section were acquired between August 7, 2010 and October 25, 2010 and cover a large part of the runs with higher luminosity for 7 Te$\,$V proton-proton collisions during 2010. The results for two fills are shown in figure \ref{fig: Cavern analysis complete fills}. The rate of registered track shapes clearly follows the instantaneous luminosity for the three different track categories. Particle types that fit into each of the three categories are detected. There are even some indications for activation of material, since enhanced rates can be seen at the end of the fills, and the falling flanks at the end of the fills are less steep than the rising flanks at the beginning. To further analyse these processes and to determine a time constant for the activations, additional measurements with more detectors are necessary.

\subsection{Comparison of registered track shapes in different layers}
\label{Comparison of registered track shapes in different layers}
\begin{figure}[htbp]
  \centering
\subfigure[Distribution of detected shapes in layers]{
	\includegraphics[width=7.2 cm]{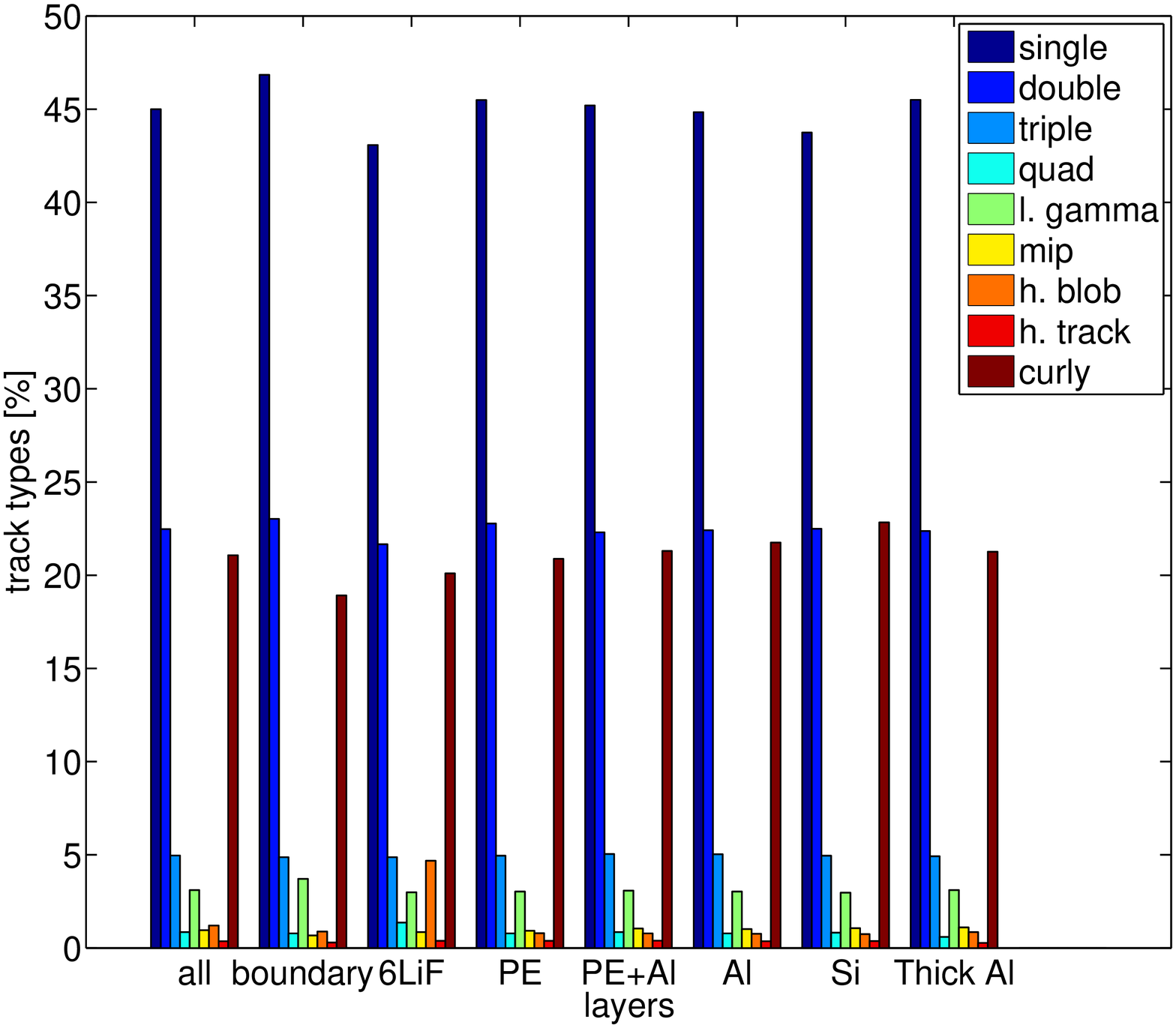}
	\label{fig: Blob types in cavern}
}
\hfill
\subfigure[Contribution of each layer to amount of detected shapes]{
	\includegraphics[width=7.2 cm]{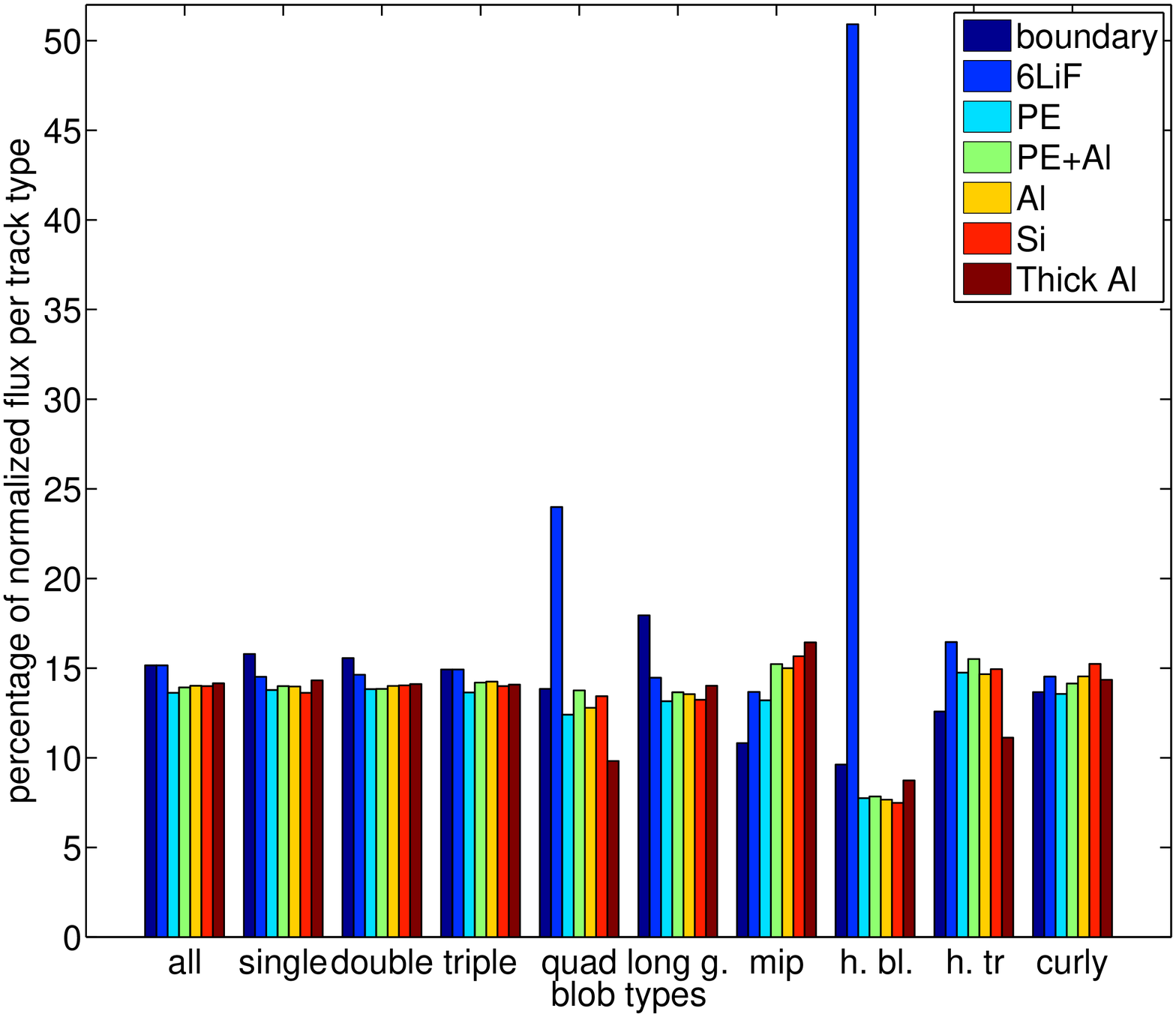}
	\label{fig: Layers in cavern}
}
\caption{Distribution of registered track shapes per layer and contribution of layers to registered amount of track shapes as measured with the Medipix detector inside the CMS cavern.}
\label{fig: Layers and blob types in cavern}
\end{figure}

Figure \ref{fig: Blob types in cavern} shows the distribution of the registered track shapes per layer. The distribution of shapes is similar in all layers with the exception of the $\mathrm{^{6}{LiF}}$ layer, which possesses a higher percentage of ``heavy blobs``. A look at the contribution of each layer to the normalized flux caused by each track type in figure \ref{fig: Layers in cavern} reveals a similar result. The $\mathrm{^{6}{LiF}}$ layer produces a far larger percentage of ``heavy blobs`` than the other layers, and its contribution to the ``quadruple hits`` is also elevated. This behaviour is likely to be caused by the conversion of incoming slow neutrons into $\mathrm{\alpha}$ particles.
The PE layer, which was expected to convert fast neutrons, does not show a significantly higher incidence of ``heavy blobs`` or ``heavy tracks``, leading to the conclusion that the conversion reaction is less efficient than assumed, or that the hit pattern is different from the expectation for this process. This needs further investigation in the future. 

The slightly higher rate of ``MIP`` shapes and the slightly lower rate of ``quadruple hits`` in the thick aluminium layer is caused by conversion and absorption in the aluminium. This indicates that the other conversion layers also contribute discriminative power between particle types since they have a different distribution of registered track shapes. However the effect was considered too subtle to be exploited at this point.

Each layer has a two pixel wide outer margin that is defined as boundary, the same applies to the complete chip. For small shapes more hits seem to occur in the boundary region than in the other layers. This is probably caused by a variety of effects. As the boundary region is less well understood, it was excluded from further analysis.

\subsection{Dependence of particle fluxes on luminosity}
\label{Influence of luminosity on particle fluxes}
\begin{figure}[htbp]
  \centering
\subfigure[Heavy-ionizing-particle-like shapes in $\mathrm{^{6}{LiF}}$ layer]{
	\includegraphics[width=7.2 cm]{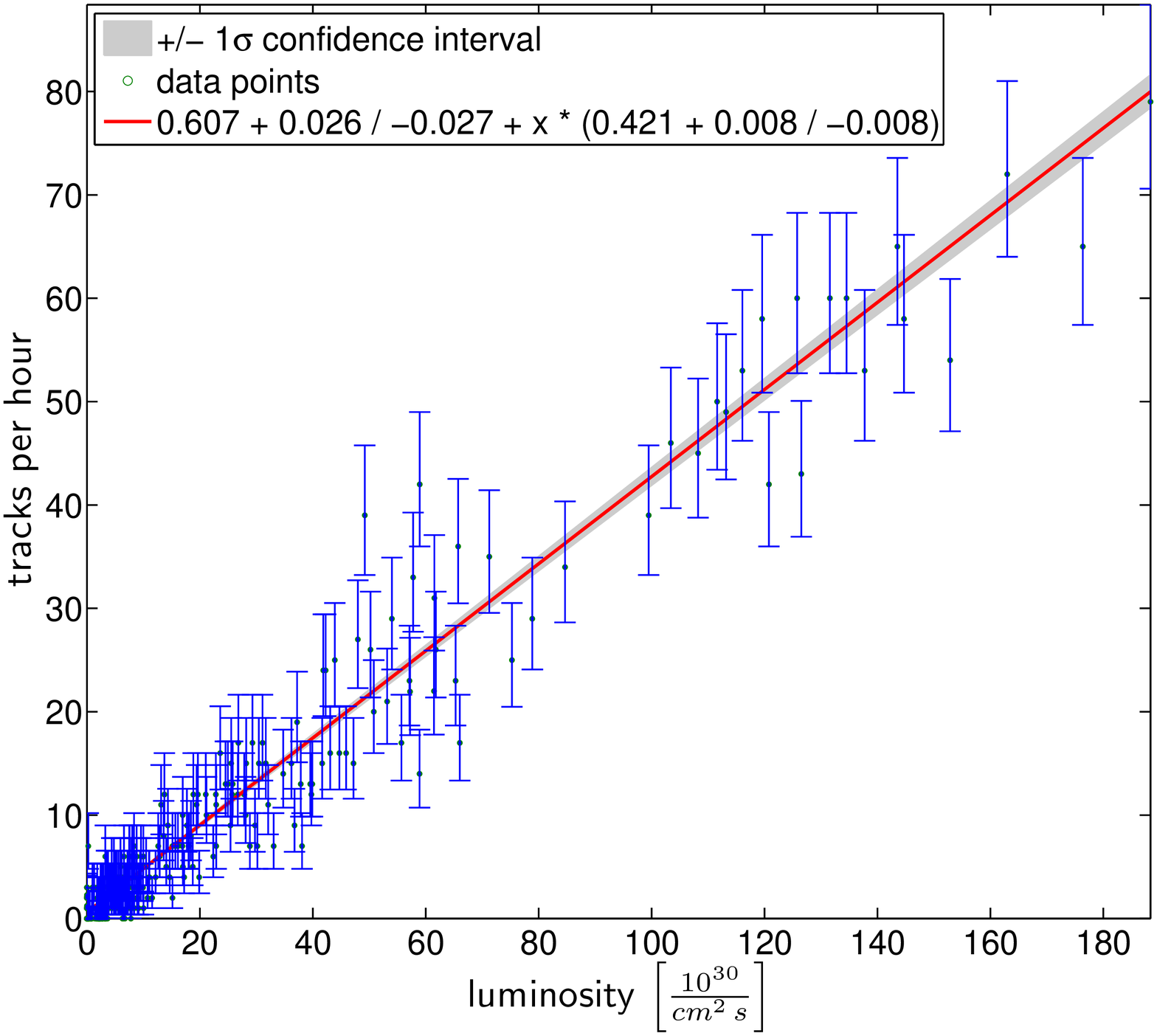}
	\label{fig: Heavy ionizing particle shapes in 6LiF layer in cavern}
}
\subfigure[Heavy-ionizing-particle-like shapes in PE layer]{
	\includegraphics[width=7.2 cm]{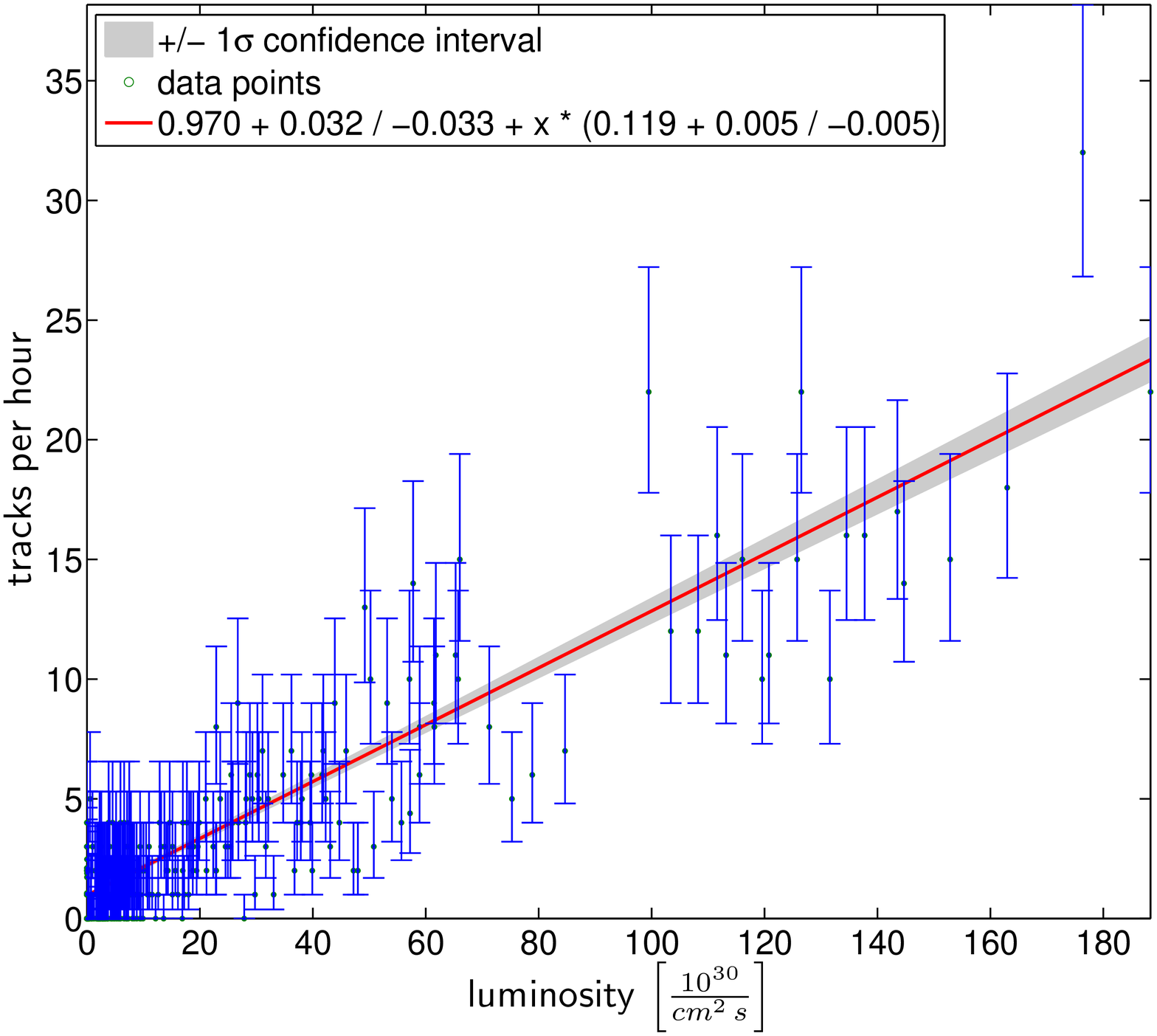}
	\label{fig: Heavy ionizing particle shapes in PE layer in cavern}
}
\\
\subfigure[Heavy-ionizing-particle-like shapes in uncovered region]{
	\includegraphics[width=7.2 cm]{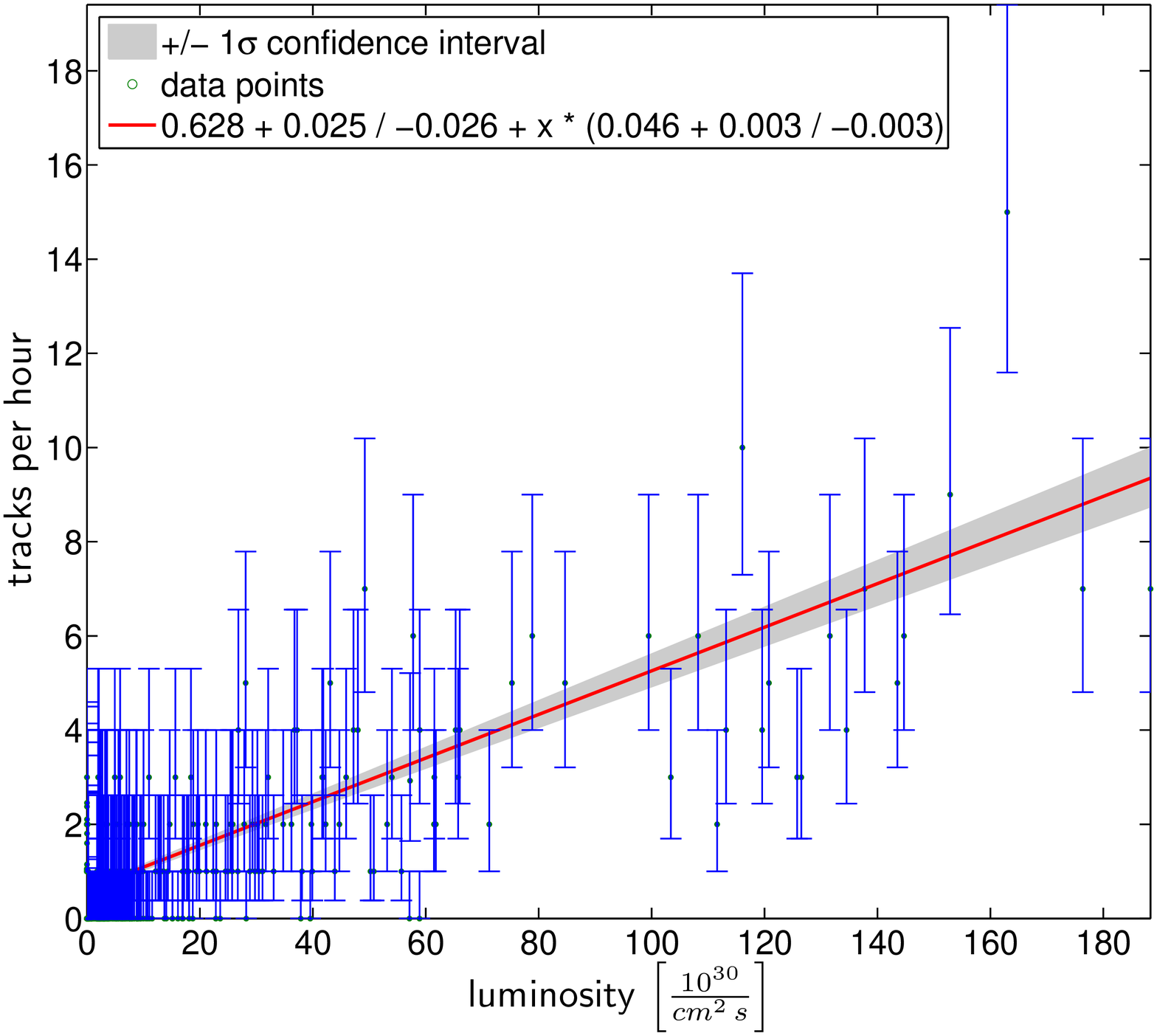}
	\label{fig: Heavy ionizing particle shapes in uncovered region in cavern}
}
\subfigure[Electron-like shapes in all layers]{
	\includegraphics[width=7.2 cm]{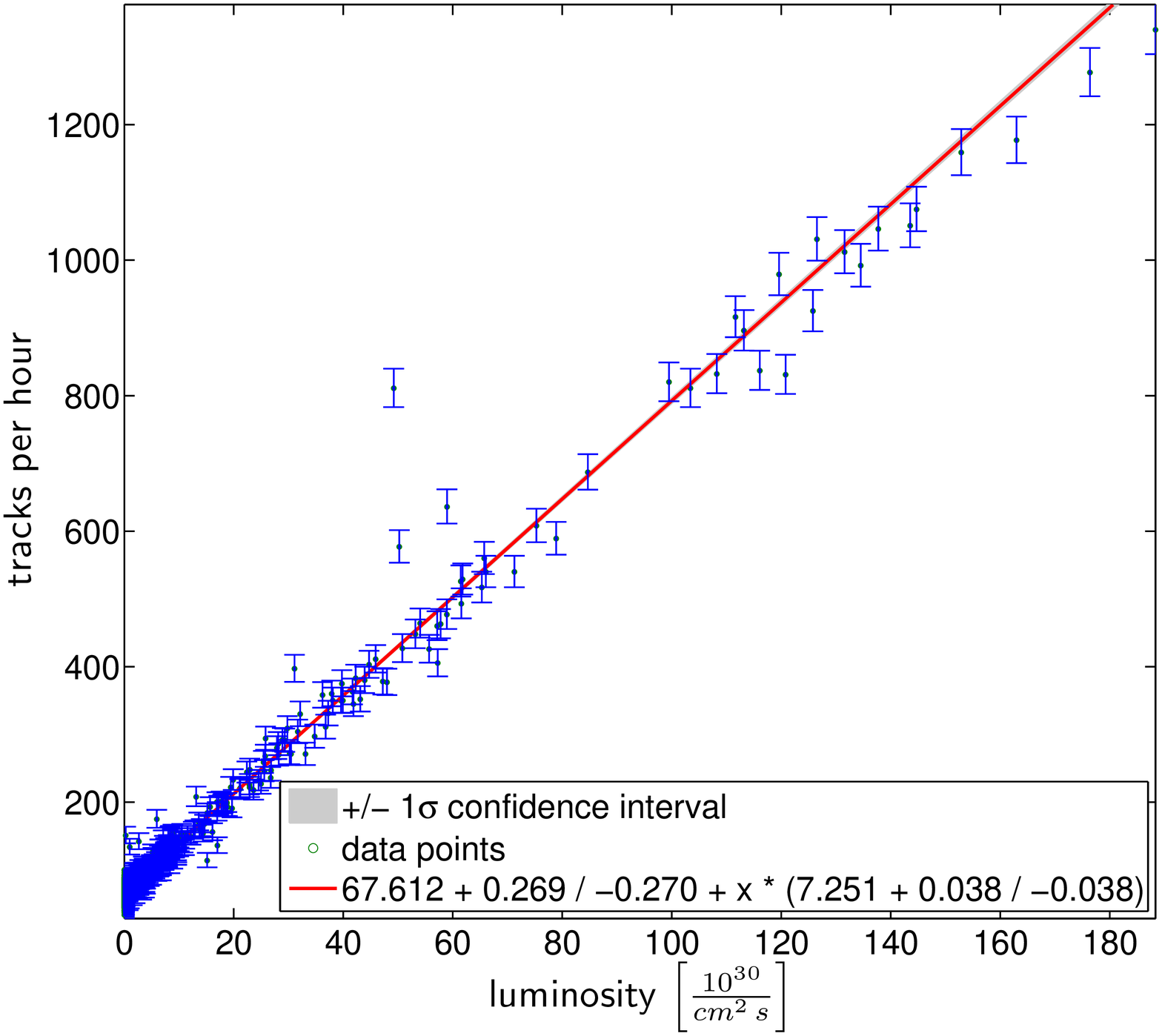}
	\label{fig: Electron shapes in all layers in cavern}
}
\\
\subfigure[Photon-like shapes in all layers]{
	\includegraphics[width=7.2 cm]{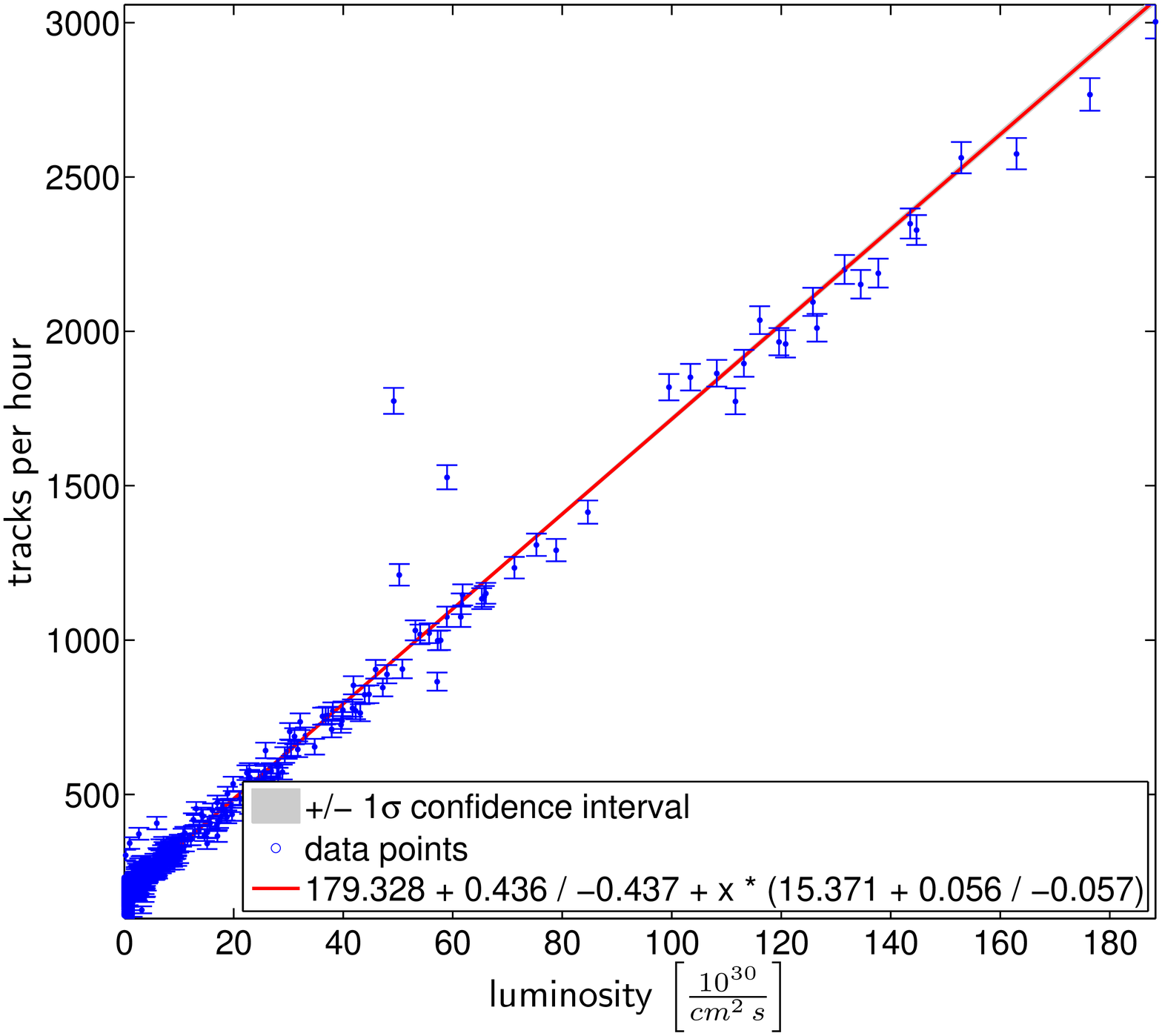}
	\label{fig: Photon shapes in all layers in cavern}
}
\subfigure[All shapes in all layers]{
	\includegraphics[width=7.2 cm]{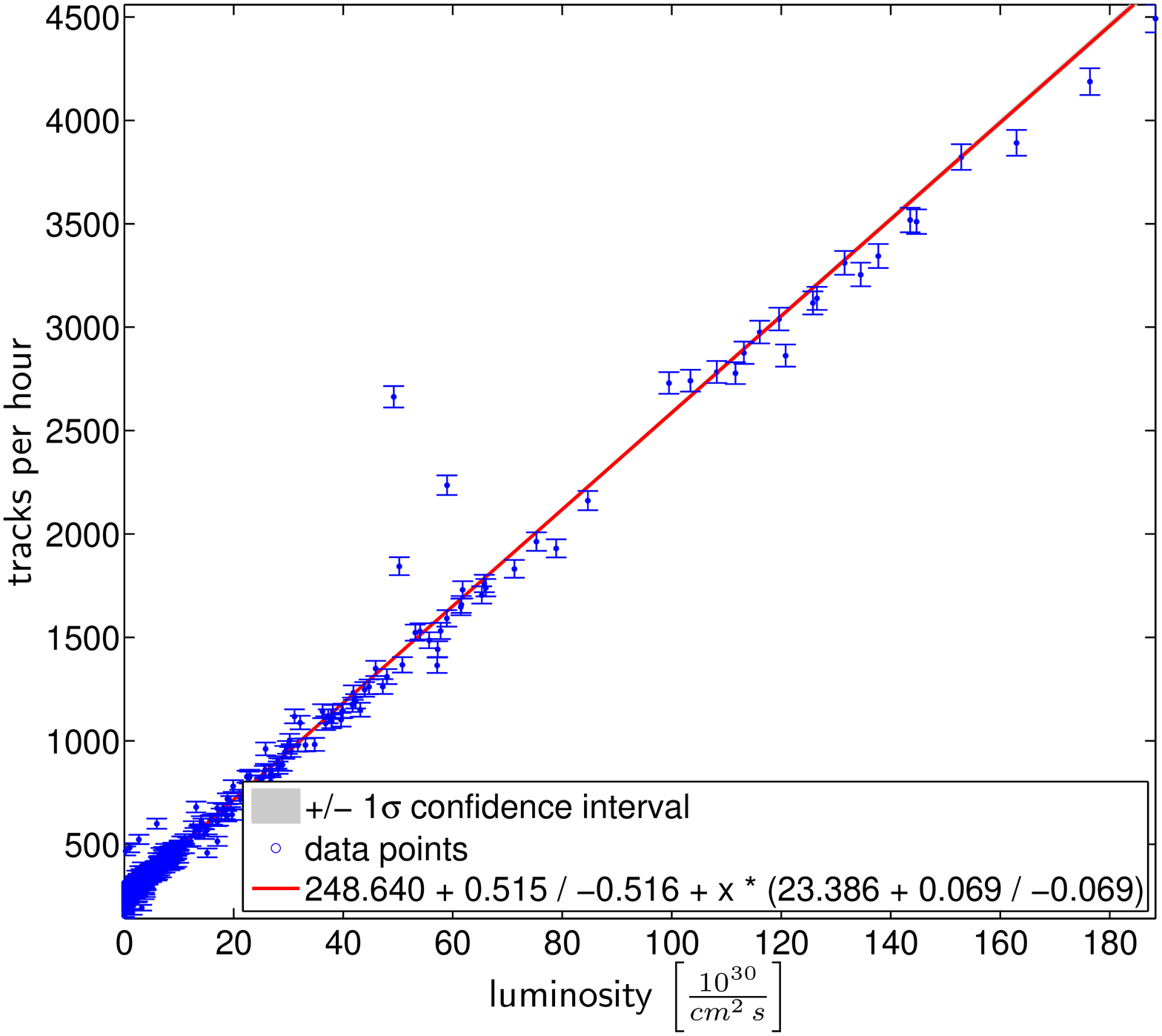}
	\label{fig: All particle shapes in all layers in cavern}
}
\caption{Detected shapes as measured with the Medipix detector inside the CMS cavern.}
\label{fig: Particles shapes in all layers in cavern}
\end{figure}

This subsection investigates the relationship between instantaneous luminosity and the number of detected particles and applies the relationship between registered track shapes and corresponding particle types summarized in table \ref{table: Blob types created by different particle types}. The shape counts are binned time-wise in one hour bins and plotted against the averaged instantaneous luminosity. Linear fits with $\chi^2$ minimization and MINOS errors \cite{MINUIT} are used to determine offset (background) and slope (rate) as shown in figure \ref{fig: Particles shapes in all layers in cavern}. At the beginning of July 2010 the instantaneous luminosity was very low and slowly increased over the month, reaching about $10^{30}\,cm^{-2}\,s^{-1}$ at the end of July. Until the end of August the peak luminosity values amounted to around $10^{31}\,cm^{-2}\,s^{-1}$. From the end of September until the end of October the instantaneous luminosity continuously increased starting with 2 $\times10^{31}\,cm^{-2}\,s^{-1}$ and reaching 2 $\times10^{32}\,cm^{-2}\,s^{-1}$ at the end of the proton run. The three exceptional data points above the fitted line in figures \ref{fig: Photon shapes in all layers in cavern} and \ref{fig: All particle shapes in all layers in cavern} as well as the four exceptional data points in figure	\ref{fig: Electron shapes in all layers in cavern} were caused by high beam losses independent of luminosity on October 24th and 25th 2010, which were also observed in other CMS beam monitoring devices.

The results from all fits are summarized in table \ref{table: Particle track fluxes}. All fits show a linear relationship between the average instantaneous luminosity and the number of registered track shapes in the cavern, but the relationship is more pronounced for the sum of all shapes, or photon-like and electron-like shapes, than for heavy-ionizing-particle-like shapes. But at least for the $\mathrm{^{6}{LiF}}$ layer a clear dependence of the number of registered track shapes per hour and the luminosity is visible in figure \ref{fig: Heavy ionizing particle shapes in 6LiF layer in cavern}. The heavy-ionizing-particle-like shapes in the PE layer in figure \ref{fig: Heavy ionizing particle shapes in PE layer in cavern} are detected in similar numbers as in the uncovered layer. Section \ref{Comparison of registered track shapes in different layers} also indicated that substantial neutron conversion only occurs via the $\mathrm{^{6}Li \, + \, n \, \rightarrow \, \alpha \, + \, ^{3}H}$ reaction in the $\mathrm{^{6}{LiF}}$ layer, whereas the additional neutron conversion via the $\mathrm{H\, +\, n\, \rightarrow \, p\, + \,n}$ reaction in the PE layer probably does not substantially add converted neutrons to the two conversion reactions $\mathrm{^{28}Si \, + \, n \, \rightarrow \, p \, + \, ^{28}Al}$ and $\mathrm{^{28}Si \, + \, n \, \rightarrow \, \alpha \, + \, ^{25}Mg}$ of the Si. A contribution of energetic protons and ions entering the silicon is also possible. According to the efficiencies determined in \cite{Greiffenberg:2009}, the number of fast neutrons converted in the PE layer should be about three times as high as the number in the uncovered region. Apparently this is not the case for the detector in the CMS cavern, suggesting that the efficiencies determined during fast neutron calibration may not be characteristic of the CMS fast neutron energy spectrum. Further investigation is needed to explain this phenomenon and to determine an appropriate conversion efficiency for fast neutrons in the CMS cavern.

\begin{table}[htbp]
\begin{footnotesize}
        \centering
        \begin{tabular}{l  c  c  c c }
     		\toprule
        Detected Shape & Layer & Layer Area & Raw Rate & Flux Shapes \\ 
& & $\left[cm^{2}\right]$ & $\left[\frac{counts}{h}/\frac{10^{30}}{cm^{2} \; s}\right]$ & $\left[\frac{counts}{cm^{2} \; s}/\frac{10^{30}}{ cm^{2}\; s}\right]$ 
\\
 \midrule  
heavy ionizing particle&$\mathrm{^{6}{LiF}}$& 0.198 & 0.421(08) &  5.91(12) $\times10^{-4}$  \\ 
electron               &$\mathrm{^{6}{LiF}}$& 0.198 & 0.870(13) & 1.22(02) $\times10^{-3}$ \\    
photon                 &$\mathrm{^{6}{LiF}}$& 0.198 & 1.802(20) &  2.53(03) $\times10^{-3}$ \\ 
all                    &$\mathrm{^{6}{LiF}}$& 0.198 & 3.091(25) &  4.33(04) $\times10^{-3}$ \\ 
 \midrule  
heavy ionizing particle& PE        & 0.484 & 0.119(05) &  6.82(28) $\times10^{-5}$ \\ 
electron               & PE        & 0.484 & 1.900(20) &  1.09(01) $\times10^{-3}$ \\ 
photon                 & PE        & 0.484 & 4.132(30) &  2.37(02) $\times10^{-3}$ \\ 
all                    & PE        & 0.484 & 6.149(36) &  3.53(02) $\times10^{-3}$ \\    
 \midrule   
heavy ionizing particle& PE+Al     & 0.338 & 0.079(04) & 6.51(33) $\times10^{-5}$ \\ 
electron               & PE+Al     & 0.338 & 1.362(17) & 1.12(01) $\times10^{-3}$ \\ 
photon                 & PE+Al     & 0.338 & 2.905(25) & 2.39(02) $\times10^{-3}$ \\ 
all                    & PE+Al     & 0.338 & 4.345(30) & 3.58(03) $\times10^{-3}$ \\  
 \midrule       
heavy ionizing particle& Al        & 0.452 & 0.098(04) & 6.00(28) $\times10^{-5}$ \\        
electron               & Al        & 0.452 & 1.874(20) & 1.15(01) $\times10^{-3}$ \\ 
photon                 & Al        & 0.452 & 3.967(29) & 2.44(02) $\times10^{-3}$ \\      
all                    & Al        & 0.452 & 5.936(35) & 3.65(02) $\times10^{-3}$ \\       
 \midrule        
heavy ionizing particle& uncovered & 0.221 & 0.046(03) & 5.83(41) $\times10^{-5}$ \\       
electron               & uncovered & 0.221 & 0.933(14) & 1.17(02) $\times10^{-3}$ \\ 
photon                 & uncovered & 0.221 & 1.881(20) & 2.36(03) $\times10^{-3}$ \\ 
all                    & uncovered & 0.221 & 2.857(24) & 3.59(03) $\times10^{-3}$ \\   
 \midrule  
heavy ionizing particle& Thick Al  & 0.078 & 0.019(02) & 6.64(79) $\times10^{-5}$ \\        
electron               & Thick Al  & 0.078 & 0.312(08) & 1.10(03) $\times10^{-3}$ \\ 
photon                 & Thick Al  & 0.078 & 0.697(12) & 2.47(04) $\times10^{-3}$ \\ 
all                    & Thick Al  & 0.078 & 1.022(15) & 3.62(05) $\times10^{-3}$ \\    
 \midrule                 
heavy ionizing particle& all       & 1.771 & 0.766(12) & 1.20(02) $\times10^{-4}$ \\ 
electron               & all       & 1.771 & 7.251(38) & 1.14(01) $\times10^{-3}$ \\ 
photon                 & all       & 1.771 & 15.375(57)& 2.41(01) $\times10^{-3}$ \\        
all                    & all       & 1.771 & 23.391(69)& 3.67(01) $\times10^{-3}$ \\ 
\bottomrule  
\end{tabular}
\caption{Track shapes of detected quanta and their rates and fluxes per layer as measured with the Medipix detector inside the CMS cavern.}                   
\label{table: Particle track fluxes}  
\end{footnotesize}
\end{table}

To derive a particle flux from the flux of the registered track shapes, the shape fluxes have to be corrected using the coefficients in table \ref{table: Blob types created by different particle types}. These coefficients represent the percentage of each particle type which is detected as a certain track given a homogeneous particle field. If one assumes that the relation between a particle and its detected track in the cavern is similar to that caused by the radioactive sources, then one arrives at the following system of equations

\begin{math}
\begin{array}{lcccccl}

  0.99\; \mathrm{\gamma} \; &+& \;0.40 \;\mathrm{e^-} \;& &                            &=&\;\mathrm{rate_{photon-like\;shapes}} \\
  0.01\; \mathrm{\gamma}\; &+& \;0.49 \;\mathrm{e^-}\;  &+& \;0.05 \;\mathrm{\alpha} \;&=& \;\mathrm{rate_{electron-like\;shapes}} \\
                           & & \;0.11 \;\mathrm{e^-}\;  &+& \;0.95 \;\mathrm{\alpha}\; &=& \;\mathrm{rate_{heavy-ionizing-particle-like\;shapes}}. 
                         
\end{array}
\end{math}  

By solving the system of equations, the ratio of particles in the cavern and the particle fluxes can be determined from the measured flux of the shapes. Everywhere besides the $\mathrm{^{6}{LiF}}$ layer one arrives at very small but negative fluxes for heavy ionizing particles. Apparently the system of equations gives a good indication regarding the ratio between detected photons and electrons, but lacks discriminatory power in determining the fast neutron flux. Taking into consideration all layers, the ratio of photons to electrons is about 40$\%$ photons and 60$\%$ electrons. After applying this ratio to the total flux of photon-like and electron-like shapes one arrives at the uncorrected fluxes for the particle types listed in table \ref{table: Particle fluxes}. The real flux is subsequently computed by dividing the uncorrected flux by the conversion efficiencies for particle types and layers. As expected in the case of relatively high energies, the resulting fluxes for electrons and photons are very similar in all layers. 

The uncorrected heavy ionizing particle flux in the aluminium layer is subtracted from the one in the $\mathrm{^{6}{LiF}}$ layer to arrive at the uncorrected thermal neutron flux. The computation of the fast neutron flux is difficult since the conversion efficiencies under the conditions in the CMS cavern are not precisely known as explained in section \ref{Particle identification and detection efficiency}. The flux caused by fast neutrons creating recoiled protons in the PE layer is lower than expected, whereas the flux due to nuclear reactions in
silicon caused by fast neutrons is higher than expected. The flux of charged hadrons in the location of the Medipix detector in the CMS cavern consists predominantly of energetic protons and could have contributed to the increased number of heavy-ionizing-like shapes in the silicon. Although the sum of both categories leads to a plausible result and is quoted here, this approach is questionable and deserves further attention in the near future. As the response of both these processes depends upon the energy spectra, it is possible that the discrepancy between the calibration spectra and the spectra in the CMS cavern is the reason for the differences observed. Nevertheless the fast neutron fluxes detected via recoiled protons and nuclear reactions in the silicon are shown here as rough estimates using the efficiencies determined during fast neutron calibration.

\begin{table}[htbp]
\begin{footnotesize}
        \centering
        \begin{tabular}{  l  c  c  c c }
           		\toprule
        Particle &  Layer  &  Uncorrected Flux & Efficiency & Flux  \\  
			&	& $\left[\frac{particles}{cm^{2} \; s}/\frac{10^{30}}{ cm^{2}\; s}\right]$ &  &
			$\left[\frac{particles}{cm^{2} \; s}/\frac{10^{30}}{ cm^{2}\; s}\right]$ 
\\    
			\midrule

neutrons (< 100 ke$\,$V)        &$\mathrm{^{6}{LiF}}$ - Al& 5.3 $\times10^{-4}$ & 5 $\times10^{-3}$ & 1.1 $\times10^{-1}$ \\ 
\midrule
neutrons (100 ke$\,$V - 20 Me$\,$V) [recoiled protons] &PE - uncovered& 9.9 $\times10^{-6}$ & 1 $\times10^{-3}$ & 9.9 $\times10^{-3}$ \\ 
neutrons (100 ke$\,$V - 20 Me$\,$V) [nuclear reactions]  &uncovered    & 5.8 $\times10^{-5}$ & 1 $\times10^{-3}$ & 5.8 $\times10^{-2}$ \\ 
neutrons (100 ke$\,$V - 20 Me$\,$V) [all]  &           &              &                    & 6.8 $\times10^{-2}$ \\ 
\midrule
electrons                      &all                 & 2.1 $\times10^{-3}$ & 1 $\times10^{-0}$ & 2.1 $\times10^{-3}$  \\ 
\midrule
photons                        &all        & 1.4 $\times10^{-3}$ & 1 $\times10^{-2}$ & 1.4 $\times10^{-1}$ \\  
\bottomrule
\end{tabular}
\caption{Particle fluxes as measured with the Medipix detector inside the CMS cavern. The row fast neutrons [all] gives the sum of the flux detected via recoiled protons in PE and via nuclear reactions in silicon, and its validity is subject to the limitations mentioned in the text. }       
\label{table: Particle fluxes}  
\end{footnotesize}
\end{table}

\clearpage

\subsection{Efficiency of shielding}

\begin{figure}[htbp]
  \centering
\subfigure[Fit for all particles in S1]{
	\includegraphics[width=7.2 cm]{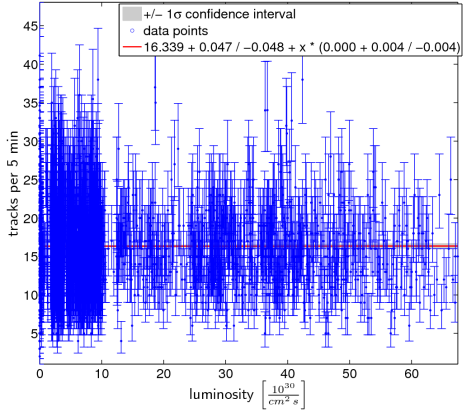}
	\label{fig: S1 fit all particles}
}
\hfill
\subfigure[Fit for all particles in CMS cavern]{
	\includegraphics[width=7.2 cm]{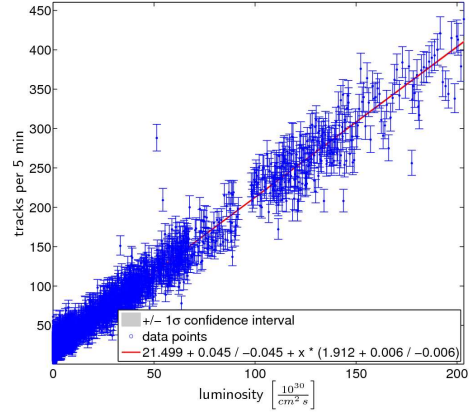}
	\label{fig: Cavern fit all particles}
}
\\
\subfigure[One fill in S1]{
	\includegraphics[width=7.2 cm]{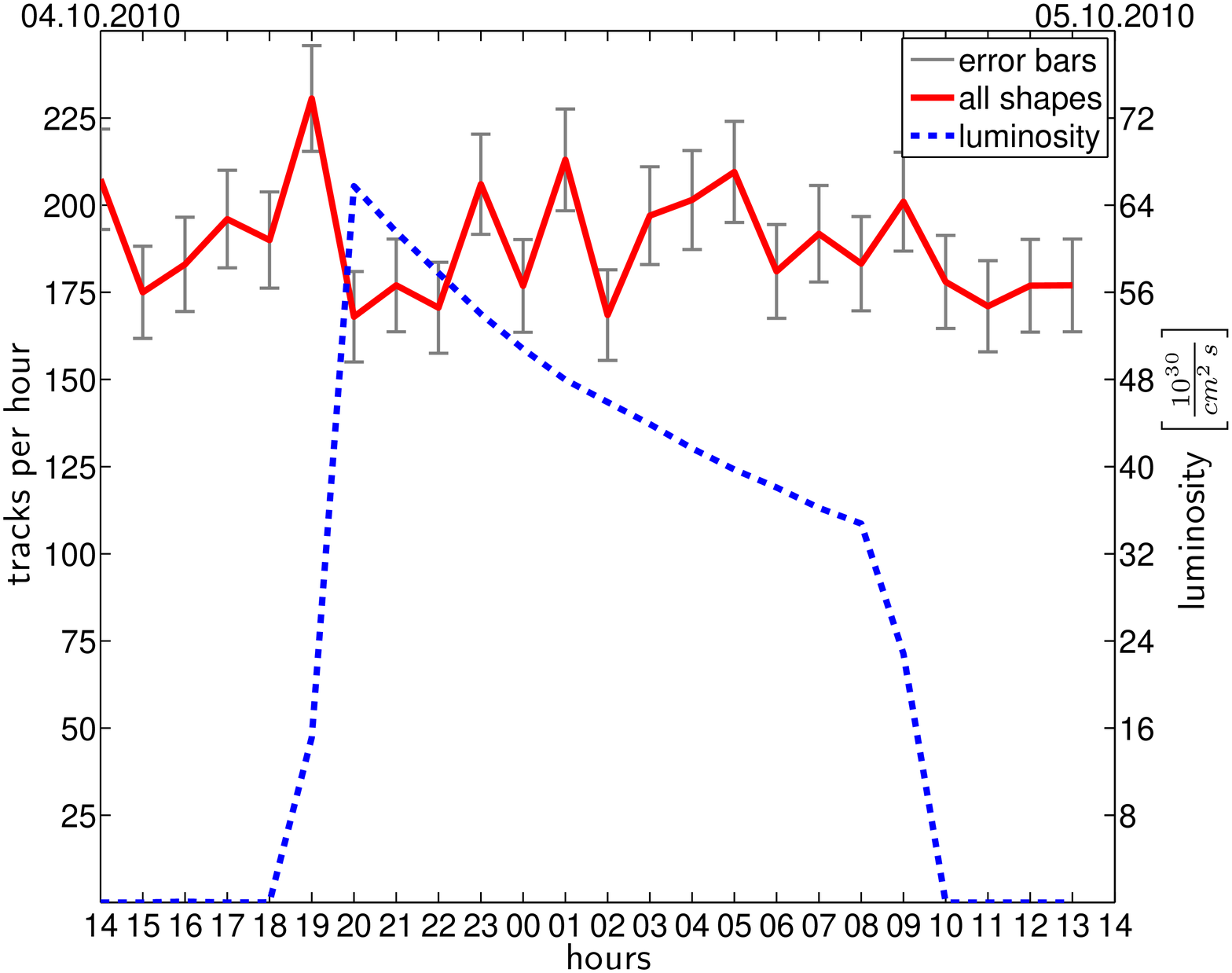}
	\label{fig: S1 one fill linear scale}
}
\hfill
\subfigure[One fill in CMS cavern with background subtracted]{
	\includegraphics[width=7.2 cm]{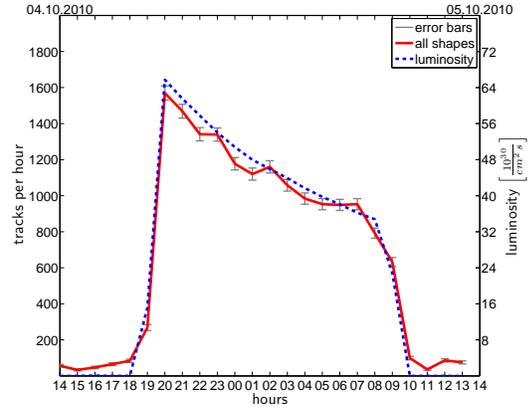}
	\label{fig: Cavern one fill linear scale}
}
\caption{Comparison of data measured with Medipix detectors in CMS cavern and S1.} 
\label{fig: Comparison S1 and cavern}
\end{figure}

The shielding of the cavern is a 7.5 m thick concrete wall, resulting in a radiation attenuation and thus a shielding efficiency in the range of 1 $\times10^5$ to 1 $\times10^6$ \cite{Huhtinen:1996uf}. An estimate for the efficiency of the shielding can be derived from the fits in figure \ref{fig: S1 fit all particles} and \ref{fig: Cavern fit all particles} as
\begin{eqnarray}
\phi_{cavern} &=& 3.6 \times10^{-3} \pm 1.1  \times10^{-5} \left[\frac{counts}{cm^{2} \; s}/\frac{10^{30}}{ cm^{2}\; s}\right] \\
\phi_{S1} &=& 0.0   \pm 7.5  \times10^{-6} \left[\frac{counts}{cm^{2} \; s}/\frac{10^{30}}{ cm^{2}\; s}\right]
\end{eqnarray}
The two sigma exclusion limit for the shielding efficiency of the cavern is therefore
\begin{equation}
  \eta =  2.4 \times10^{2}.
\end{equation}
Figures \ref{fig: S1 one fill linear scale} and \ref{fig: Cavern one fill linear scale} also demonstrate the effectiveness of the shielding. Whereas the particle flux in the cavern follows luminosity, the detector in S1 shows a constant background independent of luminosity. 

This measurement shows potential that with development and optimization these detectors can measure the efficiency of shielding arrangements.

\section{Comparison between the measured and simulated fluxes}
Taking the fluxes for thermal neutrons from the results of the $\mathrm{^{6}{LiF}}$ layer, and the sum of the fluxes for fast neutrons between 100 ke$\,$V and 20 Me$\,$V as explained in section \ref{Influence of luminosity on particle fluxes}, the complete results of the measurements are compared with the simulations in table \ref{table: Comparison of particle fluxes with simulations}. The fast neutron measurement is dependent on certain assumptions which need further investigation as detailed in section \ref{Influence of luminosity on particle fluxes}. 

All measured fluxes agree well with the simulated ones. The ratio of measured flux divided by simulated flux is within 10$\%$ of expectation for all particle types. Very fast neutron with energies over 20 Me$\,$V could not been measured, since the efficiencies of the neutron conversion layers are too low.
\begin{table}[htbp]
\begin{footnotesize}
        \centering
        \begin{tabular}{  l  c  c  c }
         \toprule
        Particle &  Measured Flux & Simulated Flux (7 Te$\,$V) & $\frac{ \mathrm{Measured\;Flux} }{ \mathrm{Simulated\;Flux} }$  \\  
			& $\left[\frac{particles}{cm^{2} \; s}/\frac{10^{30}}{ cm^{2}\; s}\right]$ &  
			$\left[\frac{particles}{cm^{2} \; s}/\frac{10^{30}}{ cm^{2}\; s}\right]$& $\left[\%\right]$ \\ 
			\midrule
neutrons (< 100 ke$\,$V)                      & 0.11   & 0.1017(14) & 108 \\ 
neutrons (100 ke$\,$V - 20 Me$\,$V)         & 0.068 & 0.0659(07) & 103 \\ 
neutrons (> 20 Me$\,$V)                       & -      & 0.0181(03) &  - \\ 
neutrons (all without neutrons > 20 Me$\,$V) & 0.178   & 0.1858(12) & 96 \\ 
charged hadrons               						&    -        & 0.000378(44) &  -  \\
electron                                  & 0.0021 & 0.0023(01) & 91 \\ 
photon                                    & 0.14   & 0.1354(19) & 103 \\ 
all (without neutrons > 20 Me$\,$V)       & 0.32   & 0.3240(23) & 99 \\

\bottomrule
\end{tabular}
\caption{Comparison of particle fluxes as measured with the Medipix detector inside the CMS cavern with FLUKA simulations.}                    
\label{table: Comparison of particle fluxes with simulations}
\end{footnotesize}
\end{table}

\section{Conclusion and outlook}
Medipix devices are installed in CMS and are working. Data from the Medipix detector in the CMS cavern show an excellent correlation between the instantaneous luminosity and the measured particle flux for all particle types. This indicates that the vast majority of particles at this point comes from interactions associated with pp-collisions. Medipix devices have the ability to discriminate between particle types, especially between neutrons, electrons and photons, which is important to determine the radiation damage of electronics in LHC. Measured fluxes show a good agreement when compared to FLUKA simulations of pp-collisions. 

The limiting factor with the present system is the understanding of the detection efficiencies for the spectra in the CMS cavern, in particular the conversion efficiencies of the neutron conversion layers. Whereas converted thermal neutrons in the $\mathrm{^{6}{LiF}}$ layer could clearly be detected, the situation is less clear in the case of fast neutrons and needs further investigation. The particle identification algorithm could be improved with tuning from the measured data.

The USB readout in combination with the USB to Ethernet extender required frequent restarts of the detector, the underlying cause of which is probably the long cable length necessary to enter the CMS cavern and the large stray magnetic field from the CMS solenoid. The next step in the extension of the current radiation measurement project in CMS is the installation of the MARS Ethernet readout for Medipix \cite{MARS}. This readout should overcome the cable length limitations of the USB readout and provide a higher frame rate, and has shown promising results in laboratory tests. A change from Medipix2-MXR to Medipix3 chips will also offer better energy resolution. 

Given the initial results from the implementation in CMS detailed in this paper, the devices have shown their eligibility as radiation monitoring devices for high energy physics mixed-field radiation environments. Looking towards the LHC upgrades in 2016, it is likely that Medipix will find wider usage as a radiation monitoring system. 

\acknowledgments
We would like to thank the Medipix2 and Medipix3 collaborations for their ingenuity and also our colleagues in the CMS Beam and Radiation Monitoring group for their advice and assistance. We are grateful to CMS Technical Coordination and the former CERN TS/LEA group for assistance in launching and sustaining the project, particularly in its early phases. We also acknowledge the contribution of the CERN, CMS, Czech Technical University in Prague and New Zealand technical teams which helped bring it to a successful conclusion. The Canterbury University group is grateful for the support by the Foundation for Research Science and Technology (FRST). R. Hall-Wilton is grateful for the support of the Israeli Technical Associates Program. We would like to thank Mark Bissell, Lukas Tlustos, Sébastien Picard and Olivia Scallon for informative discussions and advice, and Cathy Farrow for helping with the installation of the detectors.

\end{document}